\documentclass[11pt,twoside]{article}
\usepackage{asp2014}
\usepackage[normalem]{ulem}

\aspSuppressVolSlug
\resetcounters

\begin{document}

\title{Coordinating observations among ground and space-based telescopes in the multi-messenger era}

\author{Erik~Kuulkers,$^1$ Matthias Ehle,$^2$ Carlos Gabriel,$^2$ Aitor Ibarra,$^3$ Peter Kretschmar,$^2$ Bruno Mer\'in,$^2$ Jan-Uwe Ness,$^2$ Emilio Salazar,$^4$ Jes\'us Salgado,$^3$ Celia S\'anchez-Fern\'andez,$^4$ Richard Saxton$^5$ \&\ Emily M.\ Levesque$^6$}
  \affil{$^1$European Space Agency/ESTEC, Noordwijk, The Netherlands; \email{erik.kuulkers@esa.int}}
  \affil{$^2$European Space Agency/ESAC, Villanueva de la Ca\~nada, Madrid, Spain;}
  \affil{$^3$Quasar Science Resources for ESA-ESAC, Villanueva de la Ca\~nada, Madrid, Spain;}
  \affil{$^4$ATG Europe for ESA-ESAC, Villanueva de la Ca\~nada, Madrid, Spain;} 
  \affil{$^5$Telespazio-Vega UK Ltd. for ESA-ESAC, Villanueva de la Ca\~nada, Madrid, Spain;}
  \affil{$^6$University of Washington, Seattle, WA USA}

%PJT this is the first paper where I see pairs of author/affil, we can leave that, not sure if that survives the ASP style. I did edit the {} hierarchy, affil is not inside author anymore. it's parallel.
%PJT the comma goes after the name and before the $^n$

\paperauthor{Erik~Kuulkers}{erik.kuulkers@esa.int}{0000-0002-5790-7290}{European Space Agency}{ESTEC}{Noordwijk}{}{2201 AZ}{The Netherlands}

\paperauthor{Matthias Ehle}{matthias.ehle@esa.int}{}{European Space Agency}{ESAC}{Villanueva de la Ca\~nada}{Madrid}{28692}{Spain}

\paperauthor{Carlos Gabriel}{carlos.gabriel@esa.int}{}{European Space Agency}{ESAC}{Villanueva de la Ca\~nada}{Madrid}{28692}{Spain}

\paperauthor{Aitor Ibarra}{aitor.ibarra@esa.int}{0000-0003-0970-4947}{Quasar Science Resources S.L.}{ESAC}{Villanueva de la Ca\~nada}{Madrid}{28692}{Spain}

\paperauthor{Peter Kretschmar}{peter.kretschmar@esa.int}{0000-0001-9840-2048}{European Space Agency}{ESAC}{Villanueva de la Ca\~nada}{Madrid}{28692}{Spain}

\paperauthor{Bruno Mer\'in}{bruno.merin@esa.int}{}{European Space Agency}{ESAC}{Villanueva de la Ca\~nada}{Madrid}{28692}{Spain}

\paperauthor{Jan-Uwe Ness}{jan-uwe.ness@esa.int}{0000-0003-0440-7193}{European Space Agency}{ESAC}{Villanueva de la Ca\~nada}{Madrid}{28692}{Spain}

\paperauthor{Emilio Salazar}{esalazar@sciops.esa.int}{0000-0002-1331-3175}{ATG-Europe for European Space Agency}{ESAC}{Villanueva de la Ca\~nada}{Madrid}{28692}{Spain}

\paperauthor{Jes\'us Salgado}{jesus.salgado@sciops.esa.int}{}{Quasar Science Resources S.L.}{ESAC}{Villanueva de la Ca\~nada}{Madrid}{28692}{Spain}

\paperauthor{Celia S\'anchez-Fern\'andez}{celia.sanchez@sciops.esa.int}{}{ATG-Europe for European Space Agency}{ESAC}{Villanueva de la Ca\~nada}{Madrid}{28692}{Spain}

\paperauthor{Richard Saxton}{richard.saxton@sciops.esa.int}{0000-0002-4912-2477}{Telespazio-Vega UK Ltd. for ESA-ESAC}{ESAC}{Villanueva de la Ca\~nada}{Madrid}{28692}{Spain}

\paperauthor{Emily M.\ Levesque}{emsque@uw.edu}{}{Department of Astronomy}{University of Washington}{Seattle}{WA}{98195-1580}{USA}

%PJT    yes, you need lines for all of them co-authors
%\paperauthor{Sample~Author2}{Author2Email@email.edu}{ORCID_Or_Blank}{Author2 Institution}{Author2 %Department}{City}{State/Province}{Postal Code}{Country}

% leave these commented for the editors to enable them
%\aindex{Kuulkers,~E.}
%\aindex{Coauthor,~A.}          % remove and add as you need
 
 %PJT  the Aindex.py command will spit out the lines you can cut and paste in here
 
 %PJT  we were told to have a uniform style (british vs american english),so despite my background, this would force me to replace things like standardise to standardize :-(

\begin{abstract}

The emergence of time-domain multi-messenger (astro)physics requires for new, improved  ways of interchanging scheduling information, in order to allow more efficient collaborations between the various teams. Currently space- and ground-based observatories provide target visibilities and schedule information via dedicated web pages in various, (observatory-specific) formats. 
With this project we aim to: i) standardise the exchange of information about observational schedules and instrument set-ups, and ii) standardise the automation of visibility checking for multiple facilities. To meet these goals, we propose to use VO protocols (ObsTAP-like) to write the services necessary to expose these data to potential client applications and to develop 
visibility servers across the different facilities.
  
\end{abstract}

\section{Introduction}

Over the last years the scientific demands for simultaneous observations across the
electromagnetic spectrum are continuously increasing. This increase has been amplified by the
detection of non-electromagnetic events of astrophysical origin, such as high-energy
neutrino events, and, in particular, gravitational wave (GW) events. It has culminated with
the detection of prompt transient gamma-rays coincident with a GW event caused by the
merger of two neutron stars (GRB170817A/GW170817). The latter event involved many
facilities on the ground and in space and represented all currently accessible wavelengths
\citep{2017PhRvD..96l2006A}. Moreover, the transient nature of the event required fast reaction times,
in order not to miss any possible `afterglow` emission. With other large-scale facilities coming online soon which will report on transient events across the EM spectrum (e.g., LSST, SKA), efficient and fast coordination is a must in order to maximize the scientific output.

Although the process to coordinate these observations is currently cumbersome, the demand for coordinated observations is high.
For example, of the observations of ESA's space-based facilities INTEGRAL and XMM-Newton, about 10\%\ and 12\%, respectively, are coordinated with other observatories (including NuSTAR, HST, Chandra, VLT, Swift). There are nice examples of successful coordinated observations that have produced great and important science results, such as:
\begin{itemize}
    \item  Follow-up observations of the very high-energy neutrino alert on 22 September 2017 by the IceCube Neutrino Observatory, IceCube-170922A. For the first time the source of such an event was found: the event originated from a flaring gamma-ray blazar TXS 0506+056. About 20 ground- and space-based observatories were involved, with about 1010 scientists participating \citep{2018AdSpR..62.2902A}.
    \item Follow-up observations of the GW event detected by LIGO/Virgo on 17 August 2017 (GW170817). This detection has revolutionized multi-messenger astronomy, as it is the first coincident detection of a gravitational wave in electro-magnetic light, i.e., gamma-rays (GRB180817A). Subsequent observations, involving about 70 ground- and space-based observatories, and about 3680 scientists, showed it to be a kilonova, due to the merger of two neutron stars \citep{2017PhRvD..96l2006A}.
\end{itemize}

The coordination we do today, however, is not without difficulties. For example, one of the risks of these mostly ad-hoc collaborations is that the observations are not always strictly simultaneous. If time scales of variability are shorter than the degree of achieved overlap of observations, then the quality of scientific conclusions can be directly impacted.
Nowadays there are ongoing efforts (and organizations) aiming to enhance the way we coordinate  multi-wavelength and/or multi-messenger astronomy. Some of them are listed here, with no claim of completeness:
\begin{itemize}
    \item AMON -- The Astrophysical Multimessenger Observatory Network, \\ 
    see \texttt{https://www.amon.psu.edu}.
    \item SCiMMA -- Scalable Cyberinfrastructure to support Multi-messenger Astrophysics, see \texttt{https://scimma.org}.
    \item Astronomy ESFRI -- Research Infrastructure Cluster (ASTERICS),\\ see \texttt{https://www.asterics2020.eu}.
    \item SmartNet -- Simultaneous Multiwavelength Astronomy research in \linebreak[4] Transients NETwork,  
    see \texttt{https://www.isdc.unige.ch/smartnet/} \\ \citep{2017NewAR..79...26M}.
    \item DWF -- Deeper Wider Faster programme, \\
    see \texttt{http://dwfprogram.altervista.org/} 
    \citep{2018arXiv180201100A}.
    \item TOMs -- Target and Observation Managers, see \cite{2018SPIE10707E..11S}.
    \item Proposal for a multi-messenger institute, see \cite{2018arXiv180704780A}.
\end{itemize}

There are various steps in the process to improve our coordination process. These are: ``getting the alerts'', ``automated coordination'' and ``easier communication''. In the next Sections we go into the subjects in more detail. Note that we mainly concentrate on transient, sudden events, which need quick reaction. Of course, the improved ways of collaboration can be of benefit to other areas in astronomy.

\section{Getting the alerts}

First, one has to be informed that a transient event is taking place or has just happened.
Various ways to communicate all kinds of transients already exist (such as the Astronomers' Telegrams [ATels], Gamma-ray Coordinates Network [GCN] or Transient Astronomy Network [TAN] Notices, Circulars and Reports, AMON alerts, SuperNova Early Warning System [SNEWS] and Virtual Observatory Events [VOEvents], see below). Receiving notification of an event may be as simple as signing up for these existing feeds/alerts. However, still these various alerts come in various, non-standard, formats.

Moreover, the prioritization of transient follow-up observations will also be a key issue in the next decade. With facilities such as LSST and SKA coming online, as well as the increase in sensitivity of
GW and high-energy neutrino facilities, the number of transient detections with an urgent demand for follow-up will skyrocket. Managing the priority and immediacy of these triggers will become a significant challenge. In addition to prioritizing types of observations, a broader prioritization of 'categories' of follow-up observations should also be established for each observatory: 
one has to determine how urgent the follow-up of a given event is and whether other observations - including follow-up observations of another transient -- should be interrupted (for example, should a search for the optical counterpart of a GW trigger be interrupted for follow-up imaging of a nearby core-collapse supernova?).

\section{Automated coordination \&\ easier communication}

When follow-up observations need to be planned, coordination is crucial. Good coordination requires good communication tools, which must be based on standard protocols. It is key to establish good communication channels (i.e., a network) with relevant people between facilities, i.e., Principal Investigators or Project Scientists (those who make decisions about the observations) and observation planners (those that build the observing schedule). E-mail is nowadays the most-used means to communicate, but it is ad-hoc and usually addresses a selected group. One possibility is to use an open, online, messaging and collaboration tool such as Slack (or a dedicated tool like SciApp\footnote{SciApp was designed at ESA/ESAC as a collaborative application, focused to exchange information and knowledge in a specific area using modern, web mobile, technologies. It could make use of protocols (such as ObjVisTAP and ObsLocTAP described in Sec.~\ref{planning}) to gather the information from any astronomical facility and display this information in a user's friendly way. It maintains all of the discussion and results of a particular observing campaign in one easily accessible area. This application is still in a beta phase and currently on hold. See {\tt https://sciapp.esac.esa.int} for a demonstration.}). 
Users can update the facilities as well as the community, in real-time, or even in advance, of planned and/or executed follow-up observations. The public, in turn, can use the information to better optimize their planned programmes.

Another key issue is {\it rapid} response. Fast transients need fast response times (both in observations and communication). Again, with automation (such as the planning/visibility info; see below) communication becomes more efficient. Based on the available info, a decision tree (e.g., can observations be coordinated with another facility observing at a certain
wavelength?) could be used to decide on a go or no-go for follow-up observations.
A particular advantage of good communication is in designing ground-based (or space-based) observations that can complement or improve each other's observations. With a fast response time, ground-based (or space-based) observations in the immediate aftermath of a trigger can provide key initial data that can be used in planning additional observations, and ensure rapid acquisition of specific observations where ground-based coverage is equivalent or even superior.

Coordination starts with basic needs. The information that should be made accessible between observatories are of three different types:
\begin{itemize}
    \item Observing Schedule: observations that are already performed with the data already in the relevant data archives are, in most of the cases, accessible to the community. However, the information of the observations that are ongoing or the planned observations are not accessible or only accessible through in-house services (like, e.g., web pages). This information is particularly relevant to allow the follow-up of the observatories operations and plans.
    \item Target visibility: targets are not always visible for a certain observatory. In fact, the information of the periods when a certain target is observable is an output of complex calculations (sometimes geometrical, but, also related with instrument configuration or environment properties). The periods when a particular target could be observable are crucial in order to schedule a coordinated observation. In case of, e.g., variable sources, to ensure that the target is visible in parallel for a certain number of observatories in the expected source high-activity period, allows the planning of relevant science use cases that, in other way, would be impossible.
    \item Communication of changes in plans: changes in the observation planning of facilities are not easy to follow by the community as these are, in many cases, subject to a large number of factors. For example, the decision to change the plans for a particular Target of Opportunity (ToO) could be the outcome of the relevance of the ToO for this observatory, the relevance of the ongoing and near future observations, instrumental aspects, weather, etc.. However, the communication to the community of the decision of the change of the observing plan would be very interesting from the scientific point of view.
\end{itemize}

There are already some efforts by the astronomical community in place to communicate alerts and ToO follow-ups. In particular, there are initiatives from the International Virtual Observatory Alliance (IVOA) to standardize the reporting of observations from astronomical events. This standard, called VOEvent, uses VOEvent feeds to which scientists and observatories can be subscribed to receive, generate or modify notifications. VOEvent messages are XML documents that include the answers to the observation characterization:
\begin{itemize}
\item <who> - responsible (author and publisher) of the information contained in the message.
\item <how> - instrumental characterization of the observation.
\item <what> - links to the data (or measurements) associated with the observation.
\item <why> - inferences about the nature of the event.
\item <wherewhen> - description of the time and place where the event was recorded.
\end{itemize}
With this protocol (or a possible extension), the third pillar of observation coordination could be covered. 

There are no standards to ensure the information provided are offered to the community in a homogeneous way. An attempt to bring all observation planning information of several observatories together was made in a calendar format: \texttt{mySpaceCal.com}. The information displayed in that calender was obtained from the individual web pages in which observatories publish their planning information. It turned out, however, that the calender was difficult to maintain. Web page formats change without control, the metadata offered are not homogeneous (no common data model), input parameters are different per observatory and basic information to create a single client is not always present or needs further conversion. It would be better if a standard was agreed with all observatories allowing clients, such as \texttt{mySpaceCal.com}, to pull the information from any observatory participating in the consortium.

\section{Planning the observations: visibility and planning information}
\label{planning}

Some degree of coordination for GW follow-up is already de facto in place as a result of joint programmes with various observatories (e.g., HST, Chandra, XMM-Newton, Gemini, NRAO). In reality, GW follow-up will also be carried out at a large number of observatories independent of existing programmes, and the ability to coordinate effectively between these facilities is necessary in order to maximize the science.
The process of long- and short-term planning in general and coordinating observations in particular are becoming more complex in the near future. Automatic elements can make coordination more efficient by cross-correlating visibility and planning information of all
involved facilities to generate an optimized observing plan. However, as noted before, currently visibility and planning (past, current, and future) information is not available in a uniform way. 

At present there is an effort, led by ESA/ESAC, to define international standards for how observing facilities can make this information available: facilities provide two services in an agreed standard format allowing clients to make queries and receive results in a dedicated format following existing VO (Virtual Observatory) Protocols. This concept has been presented to the IVOA. The implementation of these services could commence rapidly after VO certification, and each facility could build a tool to access the information from the services of all other facilities (simple examples are given in Figs.~\ref{fig1} and \ref{fig2}). Two new protocols are now in the IVOA standardization process: Object Visibility Simple Access Protocol (ObjVisSAP, Sec.~\ref{subsec:OVSAP}) and Observation Locator Table Access Protocol (ObsLocTAP, Sec.~\ref{subsec:OLTAP}).

\articlefigure{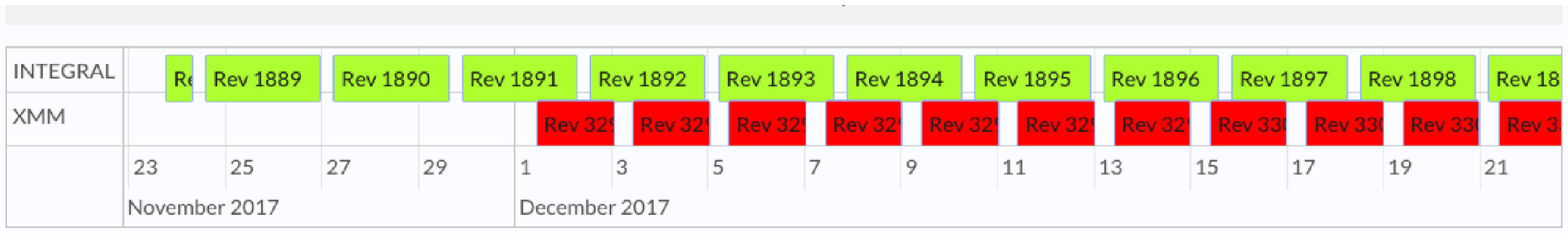}{fig1}{Simple example of visualising visibility constraints from two space-based facilities, i.e., INTEGRAL and XMM-Newton.}

\articlefigure{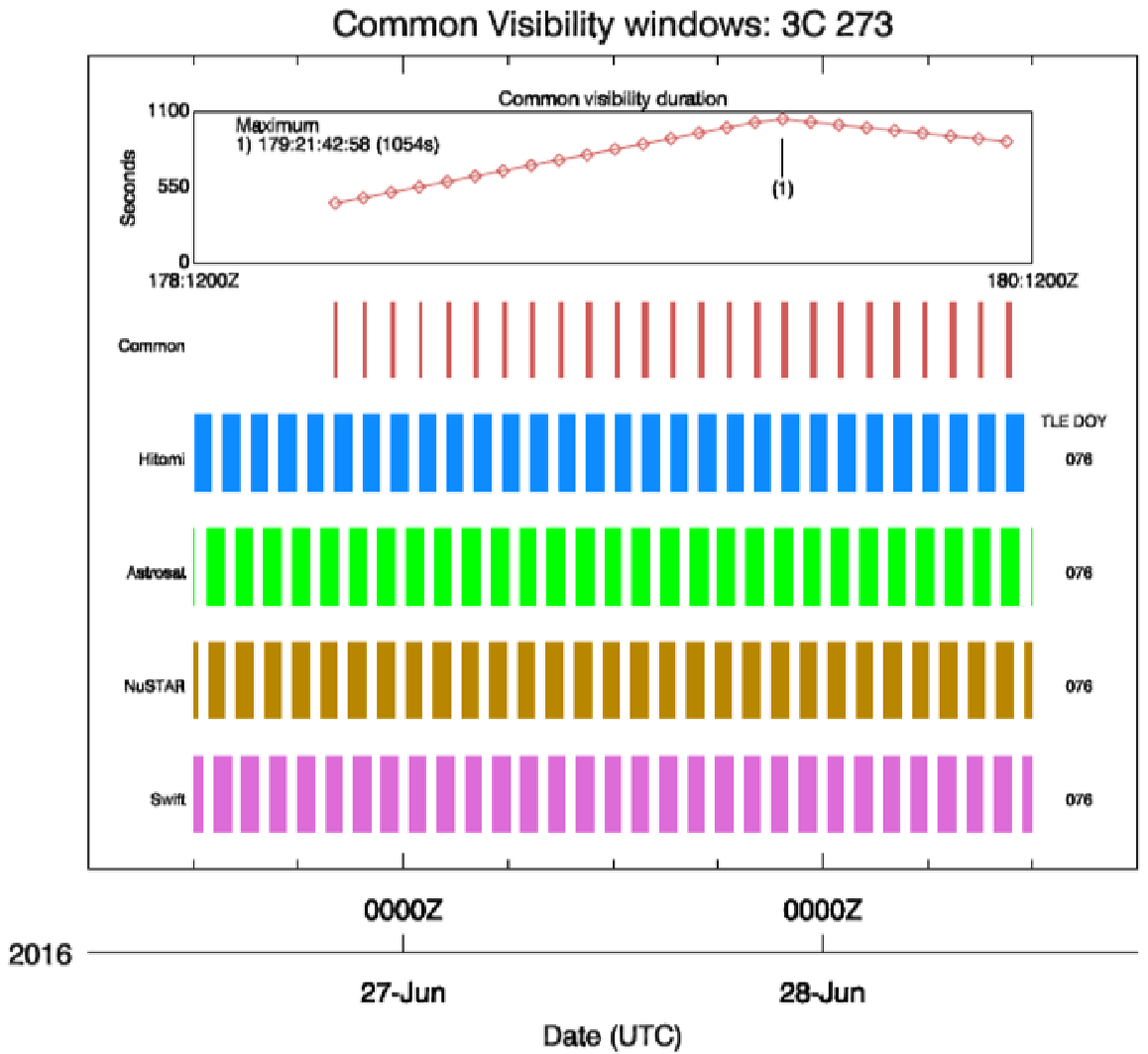}{fig2}{Visualising the optimal coverage ({\it top}) by using the target visibility information provided by various observatories.}

A workshop to discuss this effort was held at ESA/ESAC, Spain on 21 September 2018 in order to discuss the details of the VO protocols, to receive feedback on the proposed standards and to seek collaborators ready to implement prototypes and operational services. There were 29 participants at ESAC and 35 connected by video, representing more than 40 observatories at all wavelengths, as well as representatives from the GW community, and different multi-messenger/wavelength initiatives (such as SmartNet, ASTERICS). The high turnout demonstrated the broad interest in this initiative and the protocols.

\subsection{Object Visibility Simple Access Protocol}
\label{subsec:OVSAP}

The Object Visibility Simple Access Protocol (ObjVisSAP) \citep{ObjVisSAP} is a simple protocol that allows to find the periods of time when a particular astronomical target -- defined by sky coordinates -- is visible. This protocol is an IVOA S*AP protocol, defined in the following way:
\begin{itemize}
\item A specific URL for each observatory followed by a simple (standardized) parameter=value interface: the query interface is defined into the protocol allowing clients to ask for a certain target position (Right Ascencion [RA] and Declination [Dec] for a certain defined epoch). Also, a qualifier for a certain time range can be provided to constrain the query.
\item Table output response: the response of an ObjVisSAP query is an IVOA VOTable document with a set of time periods where the target can be observed by the facility.
\end{itemize}

ObjVisSAP has been kept very simple in order to facilitate its implementation by as many observatories as possible. However, although the protocol is simple, the implementation of these services by many observatories will allow access to information that is currently cumbersome to extract and allowing the preparation of coordinated proposals between many observatories. Future evolution of this protocol could cover a more complex definition of visibility (e.g., spatial coverage for a certain time), however, always as optional features.

\subsection{Observation Locator Table Access Protocol}
\label{subsec:OLTAP}

The Observation Locator Table Access Protocol (ObsLocTAP) \citep{ObsLocTAP} will allow a common interface for many astronomical facilities to publish the observation planning for current and future observations. It could also cover past observations, but, as noted above, the access to past observations is usually done via the observatories archives.
This protocol is similar to the IVOA ObsTAP protocol (for archived observations) but removing the requirements of links to the data (for obvious reasons) while adding metadata related to the instrumental configuration. Also, the protocol will allow hiding of information of the future observations whenever relevant. For example, in some cases the target to be observed is considered proprietary, so the information provided could be a time slot for a certain observation and a qualifier on the priority to indicate if this observation could be removed in case of an astronomical event.
As for ObsTAP, there are some technical details that characterise the protocol:
\begin{itemize}
    \item Protocol based on TAP: IVOA Tabular Access Protocol (TAP) \citep{2010ivoa.spec.0327D} is an IVOA protocol that publishes the information in a tabular way and provides an access to the metadata in a very close relation to a relational database. 
    \item ADQL language: instead of SQL, the language used to query a TAP service is the IVOA Astronomical Data Query Language (ADQL) \citep{2008ivoa.spec.1030O}. ADQL is similar to SQL, but removing data base dependent peculiarities of SQL flavours and adding geometrical functions that simplify queries on the sky. Other languages could be implemented in a TAP service, but ADQL is compulsory to allow the implementation of simple clients. 
    \item Data model: ObsLocTAP services have a common data model, so the same ADQL queries can be sent to the different services by a client and receive compatible/comparable metadata in return. It simplifies the implementation of clients by the requirement of mapping at server side.
\end{itemize}

\section{Conclusions}

The era of time-domain and multi-messenger astronomy (when hundreds of astrophysical
alerts happen on a daily basis) leads to the need for improved communication and efficient
managing of future transient events. New ways of communication need to be set up and
(automatic) decision trees need to be put in place, allowing to maximize the scientific
output of follow-up observations to transient events. Facilities can
enhance their science impact through participation in these processes. 
We recommend implementing a public and easy-to-use
communication system that the community can use to share information about guaranteed,
planned, and recently-executed observations. The planning information for these
observations should follow existing VO protocols. 

\acknowledgements
We thank the participants to the workshop at ESA/ESAC on 21 September 2018 for their engagement, and we hope for a fruitful cooperation and implementation of services using the discussed protocols in the near future.

\end{document}